# The Stormy Life of Galaxy Clusters: astro version[1]

Lawrence Rudnick, University of Minnesota. January 2019.

**IN THE BEGINNING**

It's 1902 and astronomer Max Wolf is confused. Scrutinizing photographic plates of the star-poor region near the north pole of the Milky Way, in the constellation Coma Berenices, he spots a dense patch of fuzzy nebulae. In his publication "The Fuzzy Objects at the Pole of the Milky Way" [1], Wolf provides careful descriptions of these objects, their sizes and morphologies -- elliptical, spiral, core brightened   He concludes with the understatement  "It would be premature to speculate on this strange result. Nonetheless, I must not miss pointing it out for general attention." (translation courtesy of Klaus Dolag, LMU and MPA).

Thirty years later, Wolf's fuzzy nebulae are shown to be "island universes," first hypothesized by the German philosopher Immanuel Kant.  They are each comparable in size to our own Milky Way galaxy, and at great distances from it.  Wolf's cluster of galaxies is today known as the Coma cluster, and its central region of massive galaxies is shown in Figure 1.

 It wasn't long before the next mystery appeared; using the Doppler shifts of galaxies in the clusters, astronomers calculated that the mass required to keep clusters like Coma gravitationally bound was ~400 times that visible in stars. In 1933, the brilliant and eccentric Swiss astronomer Fritz Zwicky attributed the required extra gravitational force to some unknown "Dunkle Materie," [2] whose nature remains a mystery to this day.

A brief glimmer of hope that the now so-called "dark matter" might have been found appeared in the late 1970s, with the discovery that clusters of galaxies were filled with an X-ray emitting plasma at a temperature of $10^{7-8}$ degrees. But estimates of its mass quickly established that it, too, fell far short of being able to bind the cluster. Our current global mass budget for clusters is shown in Box 1. Dark matter, whatever its nature, dominates the overall cluster collapse and subsequent dynamical evolution.  But a much richer story of clusters' continuing activity emerges when we examine the baryons -- not only the star-filled galaxies, but the Intra-Cluster Medium (ICM), the dominant cluster-filling plasma.

Where do clusters fit in to the overall structure of the universe?  Both computer simulations and observational studies of the three-dimensional distribution of matter show us that it is concentrated into filamentary structures that are evident approximately a billion years after the Big Bang. At the intersections of these filaments we find clusters of galaxies, with a continuing infall of dark and baryonic matter along the filaments fueling clusters' growth. Figure 1 provides a cartoon illustrating this picture.

---

[1] *Note:  This is a considerably more detailed version of the article appearing in the January, 2019 Physics Today. As per PT style, references are extremely limited.*

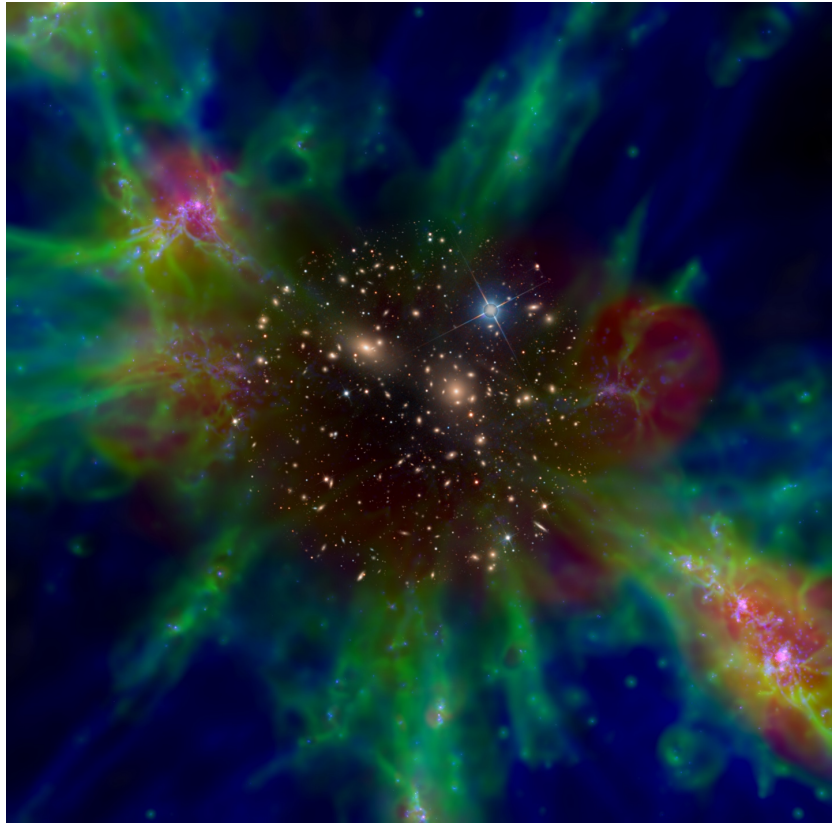

Figure 1. Illustration of clusters of galaxies forming at the intersection of large scale filaments. Optical photograph courtesy of Bob Franke, Focal Pointe Observatory showing the inner 600 kpc of the Coma Cluster. Larger scale density structure courtesy of Richard Bower, from the EAGLE simulation [3].

| BOX 1: MASSIVE GALAXY CLUSTERS: GLOBAL PROPERTIES | | | |
|---|---|---|---|
| MASS BUDGET. Total $10^{15}$ Msun; "Virial" Radius ~2 Mpc | | | |
| Dark Matter: 85% | Hot plasma (ICM) 13% | Stars & cold gas 2% | Cosmic Rays* $10^{-6}$ % |
| ICM ENERGY DENSITY BUDGET. Total $10^3$ J/m$^3$ | | | |
| Thermal 88% | Kinetic 10% | Cosmic Rays* 1% | Magnetic Fields 1% |

Notes to Box 1: These values very roughly correspond to the volume within the "virial" radius, where the mass density is >~200x the critical mass density of the universe (i.e. 200 x 10 H atoms/m$^3$ at the current epoch). Cosmic rays (CR) have energies $\gamma mc^2$, where $\gamma \gg 1$, and comprise the "relativistic plasma."

## WHO CARES ABOUT CLUSTERS?

Why are galaxy clusters worthy of our attention?  For cosmologists, clusters provide unique signatures of the evolution of the universe.  The overall growth of structure in the universe is regulated by a few key parameters, such as the fraction of mass in baryons compared to dark matter and the overall pressure and nature of dark energy.   The number density of the most massive clusters and their distribution in redshift (time) provide very sensitive probes of such parameters.  And for dynamicists pondering how gravitationally bound structures form and come into equilibrium, clusters offer a critical laboratory.  They are the largest bound structures that currently exist, illuminating the transition from the linear infall regime to the non-linear relaxation stage.   The diffuse cluster plasma offers even more rich opportunities for astrophysicists.  It allows us to explore a system with a full census of its baryons, unlike the situation for galaxies and less massive groups of galaxies.   Therefore, for issues such as the evolution of galaxies in clusters, we can trace the effects of shocks and ram pressure on the formations of stars and galaxies from the observed intracluster medium, without worrying about what we've missed.  And this "collisionless" high-beta (low magnetic field) plasma itself offers bountiful opportunities (to put a positive spin on our ignorance!) to explore interesting physics, e.g., turbulent cascades, ion-electron equilibrium, amplification of magnetic fields and the coupling of cosmic rays to MHD structures. Interested readers are commended to the review by Andrei Bykov et al. [4]

## WEATHER VANES for the INTRACLUSTER MEDIUM

There's one more item of business before we can explore the dynamics of the intracluster medium -- the radiative diagnostics.   Unfortunately, we must limit ourselves here to exploring the thermal and relativistic plasmas, with exciting new information from the studies of atoms and molecules in the relatively cooler gas near the massive central cluster galaxies left for another place and time.

Box 2 summarizes the main radiative processes for these two plasmas.  Each mechanism both scales with overall cluster parameters and produces a distinct spectral shape (e.g., $T^{-0.5} \exp[-h\nu/kT]$ for thermal bremsstrahlung and $\nu^{-\alpha}$ for synchrotron radiation, where $\nu$ is the observing frequency).  These spectral distributions depend both on the single-particle radiation spectra and the energy distributions of the radiating particles -- Maxwellian for the thermal plasma and a power law for the cosmic rays.  Additional information can be derived from these spectral distributions, such as the temperature of the X-ray emitting gas.  Since multiple mechanisms depend on the same cluster physical quantities, we can sometimes combine different measurements to break the degeneracies.

| Box 2. ICM: Scaling of Radiative Diagnostics | | |
|---|---|---|
| | Thermal plasma | Cosmic Rays |
| Thermal plasma | Bremsstrahlung (X-ray) $n_e n_p T^{0.5}$ H- and He-like line emission | Pion production ($\gamma$-ray)`` $n_p n_{CRp}$ |
| Magnetic field | Faraday rotation (radio) $n_e B_{\parallel}$ | Synchrotron (radio) $n_{CRe} B^2$ |
| CMB | (t) S-Z effect (mm) $n_e T$ | Inverse Compton (X-ray) $\gamma^2 n_{CRe} T_{CMB}^4$ |

Notes: Emissivity scalings are "bolometric", integrated over all frequencies, without showing the spectral dependencies. $n_e$, $n_p$, $n_{CRe}$ and $n_{CRp}$ are the densities of electrons, protons in the thermal and relativistic plasmas, respectively. T is the (average) temperature of the thermal plasma and $T_{CMB}$ is the temperature of the CMB at the redshift of the cluster. B is the characteristic strength of the cluster field, while $B_{\parallel}$ is the vector average of the magnetic field component along the line of sight. $\gamma$ is the relativistic factor, typically around 1000 for CR electrons observed at radio frequencies.

Basic information about clusters comes from spatially integrated measures of these radiative diagnostics, and how they scale with physical cluster properties such as mass or redshift. Much more value is added when the radiation can be mapped across the cluster. A rich literature of cluster maps in X-ray and radio emission exists, and observations of the S-Z effect (see below) are rapidly increasing. But despite heroic efforts, and great expectations for the Fermi satellite, there is currently no clear detection of cluster $\gamma$-rays, which would provide our only window on the CR proton distribution. Similarly, there is not yet a clear detection of the inverse Compton radiation from CR electrons, which could be combined with the synchrotron emissivities to measure the strength of the cluster magnetic fields.

Although we don't have magnetic probes flying through the cluster plasma, the magnetic fields reveal themselves through the synchrotron radiation from CR electrons, and the Faraday rotation of linearly polarized radio emission through the magnetized ICM. The strength and structure of the magnetic fields can illuminate the characteristic turbulent scales in the hot plasma to which they are coupled. Although their energy densities/pressures are low compared to the ICM thermal pressures, the magnetic fields influence the transport of heat in the ICM and the energization of CRs.

One radiation mechanism has a curious spectral signature. It is the inverse Compton scattering of CMB photons to slightly higher energies by the hot ICM electrons. This mechanism is known as the thermal (t) S-Z effect, named for Rashid Sunyaev and Ya. B. Zel'dovich, who first described it in 1972 [5] (and sent the author off on a not sufficiently sensitive search for it at that time).  In the Rayleigh-Jeans part of the spectrum (wavelengths longward of ~1.4mm), the shift of the spectrum to higher energies creates a **lower** brightness compared to looking off the cluster.  Subtracting *on-off* measurements therefore yields a **negative** signal. By contrast, a positive *on-off* signal is seen at shorter wavelengths.  Today, mapping the (t)S-Z effect is a worldwide industry, producing line-of-sight integrated ICM pressure measurements that can be combined with X-ray data to derive robust physical ICM parameters, identify shocks, etc.

Our current characterization of the ICM based on these diagnostics is shown in Box 3. The thermal plasma is quite different than terrestrial and even other astrophysical plasmas.   The extremely large number of electrons within the Debye sphere (~$10^{14}$) makes the ICM one of the most perfect ``collisionless" plasmas in the universe. The effective scattering length is set by plasma waves, for which there are an endless stream of possibilities, perhaps down to scales as small as the ion gyroradius.

| Box 3. ICM CHARACTERISTIC PARAMETERS | | | | |
|---|---|---|---|---|
| **Temperature** $10^{7\text{-}8}$K | **Density** $10^{2\text{-}4}$ m$^{-3}$ | **Sound speed** ~1000 km/s | **Particle mfp** ~ few kpc | **Plasma scattering length** >~ $10^8$ m |
| **Magnetic field** <0.1-1 nT | **Plasma Beta** ~100 | **Alfven speed** 20-100 km/s | **Magnetic field typical scale** ~10s kpc | **CR Diffusion time, ~Mpc** $10^{10.5}$ y @1GeV |

Notes:  All cluster properties are functions of distance from the center, not shown here.

Finally, there are literal weather vanes in clusters, where low density, usually bipolar jets of plasma are ejected from supermassive black holes at the centers of some galaxies, the so-called Active Galactic Nuclei (AGN).  These jets of plasma are deflected and distorted in their relative motion through the ICM - some examples will be seen below.

## GENTLE STORMS

Clusters have been growing over the last 10 billion years through the continuing infall of material, mostly clumps of dark matter, primarily along the filamentary structures (Figure 1).  In some clumps, the accompanying baryonic matter has cooled sufficiently for stars and galaxies to form. Once these galaxies reach the denser environment of the cluster cores, they are scattered by local irregularities in the gravitational potential.  Their velocities begin to isotropize and develop a randomized velocity structure typical of systems in virial equilibrium.  The "virial radius" for rich clusters, within which there has been sufficient time to approach this state, is typically about 2 Mpc for mass >$10^{15}$ Msun clusters, and virial velocities are of order 1000 km/s.

When the infalling baryons are still in a diffuse state, they experience a quite different fate.  An "accretion shock," seen in simulations but not yet in observations, forms about 2 times further out than the virial radius from the cluster center. As the infalling diffuse material encounters the shock, their infall energy is almost entirely converted into heat. This brings the gas to an equivalent "virial temperature" of approximately $10^8$K, thus forming the ICM plasma. As the stars within cluster galaxies continue to evolve, winds and shocks transport material into the ICM, enriching it with what astronomers call "metals," the elements from C to Fe that were produced in stellar nucleosynthesis.

Given sufficient time without major perturbations, clusters can take on a "relaxed" state, with the infall of gas and galaxies contributing to the growth of a massive central galaxy, $10^{13}$ Msun. The radiative cooling time in the inner regions is shorter than the age of the cluster, and the temperatures can drop towards the cluster center. Such relaxed clusters have high X-ray luminosities and are prominent in cluster surveys. They are recognized by their bright, central cores, their regular velocity distributions, and their generally symmetric appearance. This apparent relaxation is deceiving.

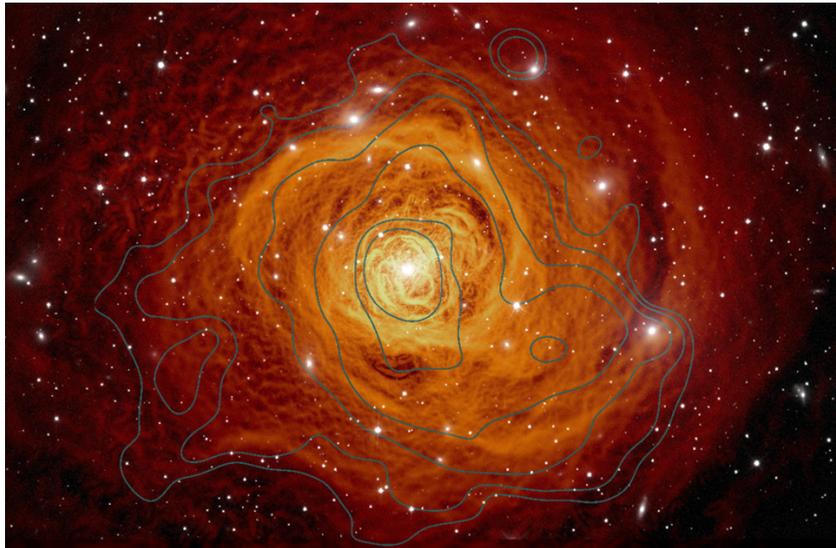

Figure 2. Inner 400 kpc of Perseus cluster in X-rays, filtered to show fine structure, courtesy of Stephen Walker, NASA Goddard, superposed with r-band image from the Sloan Digital Sky Survey and radio contours of the "mini-halo," courtesy of Marie-Lou Gendron-Marsolais, University of Montreal.

Occasionally, the infalling clumps of gas and baryons are a substantial fraction of the cluster mass. In that case, there is little prompt heating of the infalling baryons and a "cold" (i.e., only $\sim 10^7$ K!) clump of gas can penetrate deep into the cluster. Many such structures are identified by their bright appearance and sharp edges. They are distinguished from shocks because they are in pressure equilibrium with the surrounding hot ICM. While remaining cold, the clump may also generate a bow shock and heat the surrounding medium. If the intruder is massive enough, it can even set off "sloshing" motions in the ICM baryons, as they are displaced from, and then restored by, the underlying dark matter core. In these cases, spiral waves can be set up, as can be dramatically seen in Figure 2. Here, a highly enhanced X-ray image of the nearby Perseus cluster of galaxies reveals the effects of a likely off-center encounter with a $10^{14}$ Msun clump approximately 2.5 billion years ago. In this case, some of the energy has also been transferred into the cosmic rays, leading to a so-called synchrotron "mini-halo" shown in contours. The sloshing motions can help mix the cooler and hotter gas, redistribute metals throughout the cluster, and induce turbulence that eventually dissipates the energy into heat. Buried deep in the center is also an AGN associated with the bright central galaxy NGC 1275. Its ejections evacuate bubbles in the X-ray gas, creating dark patches.

**VIOLENT STORMS**

Inevitably, in this process of infall along filaments, there will sometimes be more dramatic collisions between massive clusters of comparable sizes, such as illustrated in Figure 3. This is the well-known "Bullet cluster" where the broght bow-shaped structure is a pressure-matched discontinuity. The brighter regions correspond to the cool, dense remnants of the original clusters. Beyond these are the fainter, but hotter regions heated by shocks formed in the >$10^{57}$ Joules encounter. In this spectacular collision, the cluster galaxies and dark matter have low enough cross sections that they passed right through each other, while the diffuse plasma bears the brunt of energy dissipation.

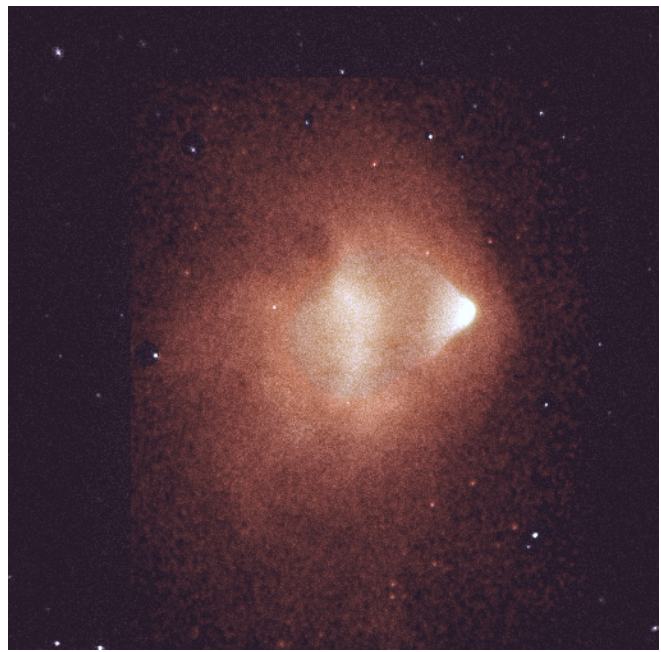

Figure 3. X-ray emission from the Bullet Cluster, courtesy M. Markevitch.

Such shocks and the accompanying post-shock turbulence can generate a rich variety of phenomena in the ICM. A spectacular example is seen in Abell 2256, (Figure 4), the ongoing merger of two or three massive clusters. Temperature maps of the X-ray emitting ICM show several pressure-matched discontinuities, different in shape but similar in physical properties, to the "bullet" shown above. Coincident with the X-ray emission is a benign looking white contour, enclosing a region of diffuse, low brightness radio synchrotron emission, which we call a "radio halo." It reveals that the smooth-looking X-ray emission is deceptive, because the ICM must be highly turbulent in order to accelerate the cosmic ray electrons responsible for the radio emission. At observed frequencies around 1 GHz, these ~1 GeV electrons will radiate away their energy in ~$10^8$ years, which means that they cannot move far from their energization sites. CR electron acceleration must therefore be happening throughout the cluster -- and that puts the burden on otherwise invisible turbulence in the ICM.

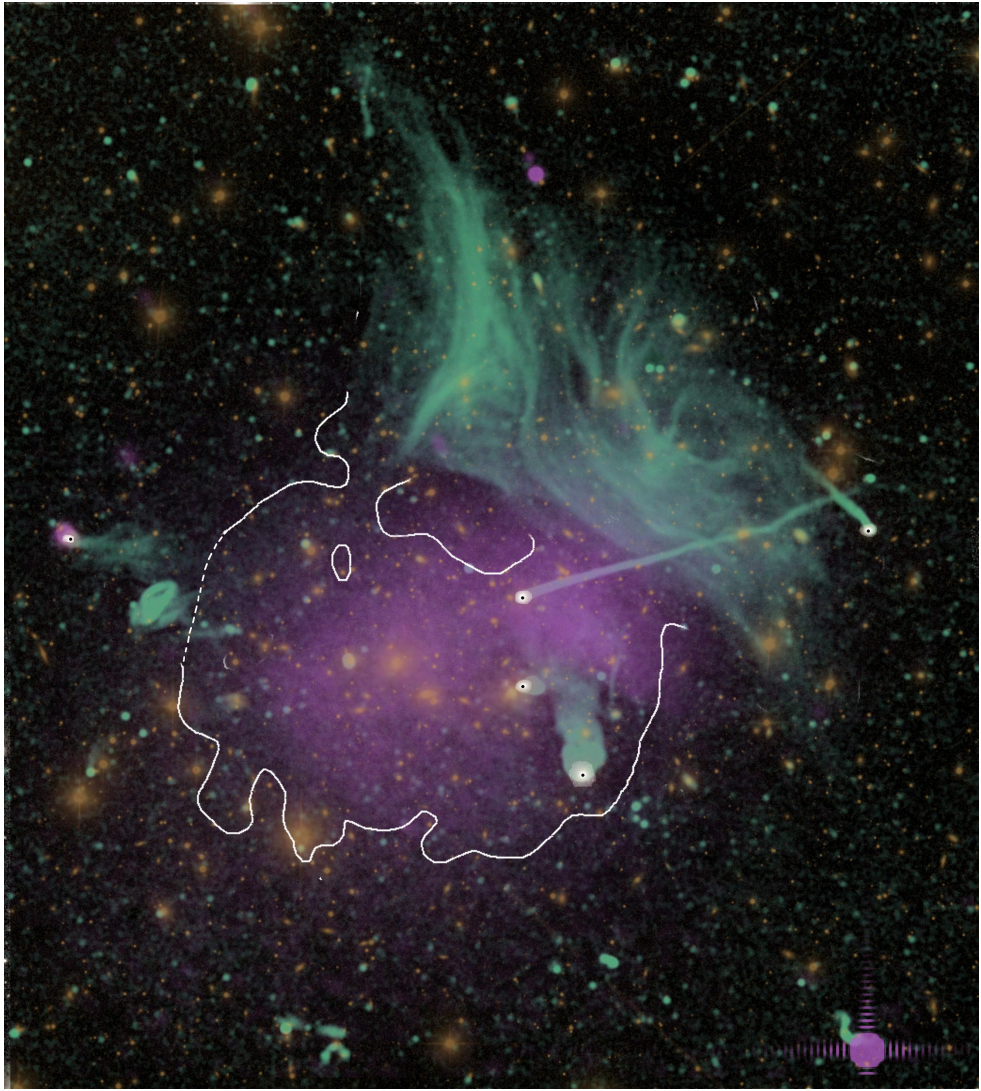

Figure 4. Abell 2256, in the 2-10 keV X-ray band (purple), near infrared band (orange) and radio synchrotron (green, [6]) The white contour indicates the extent of the diffuse radio "halo". The white patches with dark spots indicate a few of the cluster galaxies that host a radio structure bent back by relative motion through the ICM.

In Figure 4, the large filamentary radio structure in the upper right is assumed to be associated with shocks generated in the cluster collision, although in this case, the X-ray shock structure has yet to reveal its presence. Such synchrotron features are often found on the periphery of cluster X-ray emission and are termed "radio relics" for historical reasons. Individual galaxies in the clusters are themselves also sources of radio emission, and their swept back appearance indicates relative motion through the ICM. In some cases, sharp deflections in these radio tails, or curious features such as the ring-like structure in the left of Figure 4 can point to otherwise invisible flows in the ICM   Such indications of "weather" in the environment surrounding radio galaxies are spotted occasionally in cluster mapping projects;  an emerging source of such critical probes are the sharp-eyed "citizen scientists" participating in projects such as Radio Galaxy Zoo.

The observed and inferred shocks and turbulence in the ICM, and features such as radio halos and relics appear almost exclusively in clusters that have recently experienced major mergers. But what is the mechanism by which they transfer energy to a small and highly energetic population of CRs?  The answer lies in the small-scales of the turbulence, where the repeated scattering of electrons by self-excited and other MHD waves slowly restores their radiated away energy.  Simultaneously, the magnetic fields required for synchrotron radiation are seeded by processes currently unknown, and then amplified largely through stretching, as they are sheared in the turbulent ICM.   Approaching such a complex set of interconnected processes requires numerical simulations, such as illustrated in Figure 5.

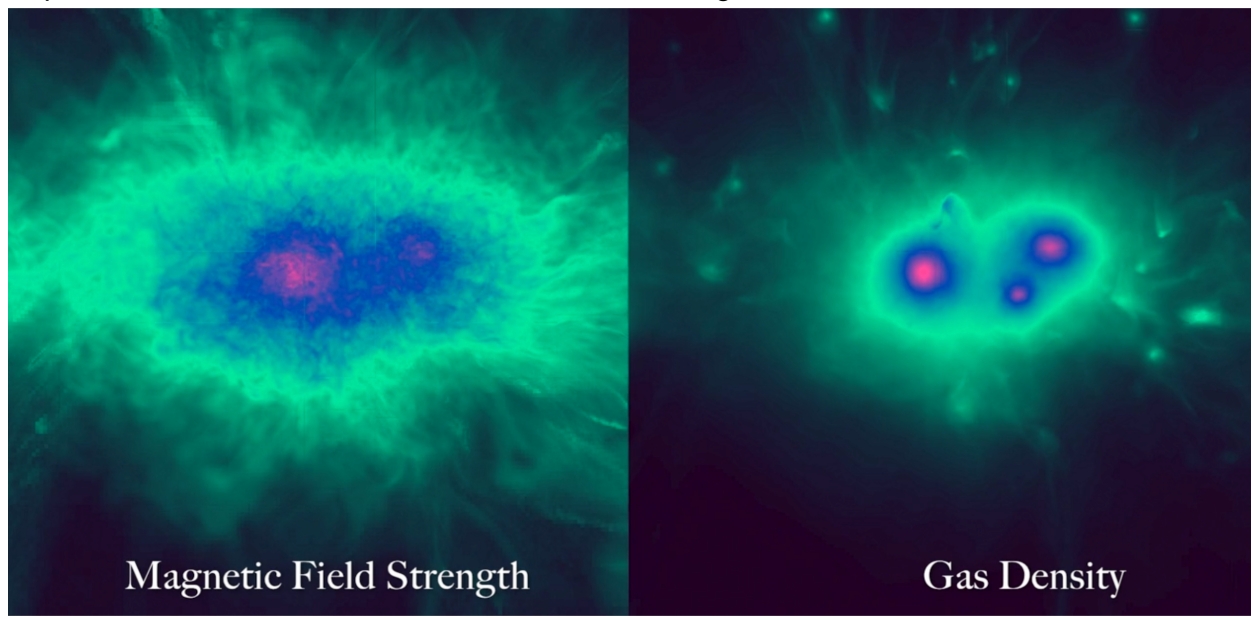

Figure 5.   Numerical MHD simulations of merging clusters showing widespread amplification of the magnetic field. (Courtesy of F. Vazza, University of Bologna, see https://vimeo.com/289054145)

Such simulations, with increasing support from both radio and X-ray observations, indicate that the turbulence and its consequences are relatively more important in the outer, lower density regions, as opposed to cluster cores where the thermal energy is highly dominant.  Extending such simulations to predict the distribution of synchrotron radiation requires understanding the microphysics of cosmic ray acceleration.  In particular, we need to know how the turbulent energy is dissipated on the smallest scales, far below the resolution of cluster-wide numerical simulations.  Processes such as the relative roles of Alfvenic, compressive, solenoidal turbulence and CR self-excited modes are under intense study. Interested readers are commended to the review article by Gianfranco Brunetti and Tom Jones [7].

Much of the ICM turbulence is likely generated downstream from shocks and gas perturbations that are produced during the merger process. In numerical simulations, a dominant pair of opposite moving shocks often forms. When these reach the peripheries of the merging cluster they become separated from the rest of the cluster emission and become more accessible to observations. One such shock, dubbed the "Toothbrush" because of its morphology, is shown in Figure 6. Here we see the radio emission generated immediately behind the shock, and a profile plot of the corresponding X-ray brightness.

The X-ray and radio emission each give us a way to estimate the shock Mach number, a key parameter in determining both what has taken place during the merger and how the ICM will be affected. In the X-ray, the Rankine-Hugoniot jump conditions can be applied pre- and post-shock to deprojected values of either the ICM density and temperature, or preferably both. In the case shown here, the X-ray shock is quite modest, trans-sonic, with a Mach number of ~1.3. Projection and smearing effects often lead to a factor of 2 uncertainty in the actual amount of compression in such shocks.

The radio emission provides an alternative insight into the shock structure. If the enhanced radio emissivity post-shock arises from diffusive shock (or Fermi type I) acceleration, then the logarithmic slope of the synchrotron radiation spectrum will reach a steady-state value that depends on the Mach number. Observations of the Toothbrush reveal that instead of the low X-ray derived Mach number, the synchrotron derived value is M~3.5 . To first order, it's remarkable that these simple first-order theories with very different underlying physics lead to similarly low Mach numbers. At second order, the apparent discrepancy between X-ray- and radio- derived values have led us to look deeper into our models.

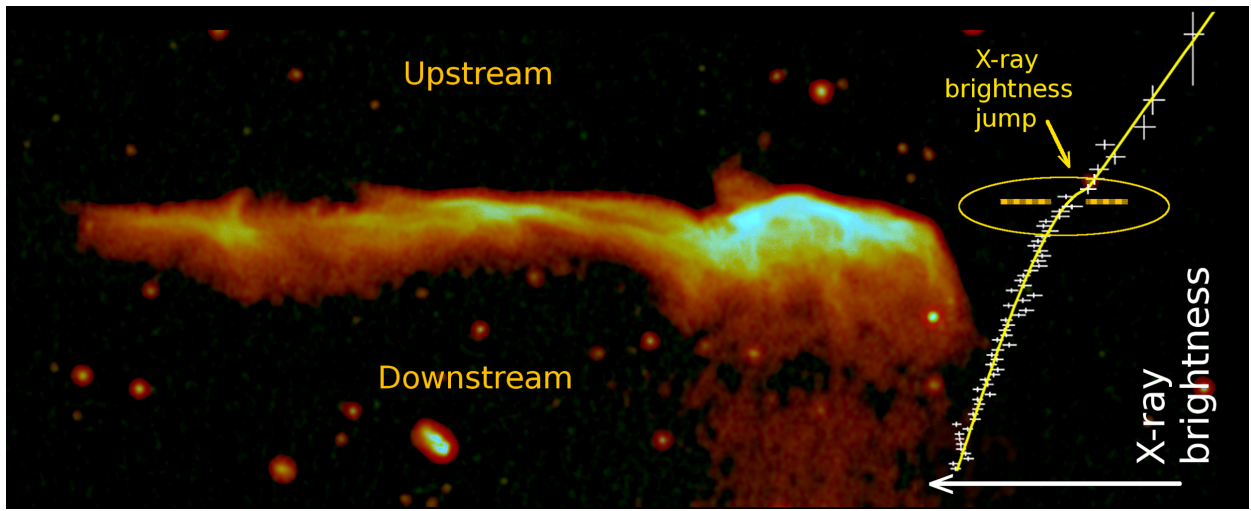

Figure 6. Radio emission from the Toothbrush Relic and its approximate correspondence with the X-ray brightness profile. See [8] and references therein.

One additional consideration is whether all of the radiating CR electrons are generated at the shock, or whether some actually come from a pre-existing population left behind by cluster radio galaxies. This latter possibility has the extra advantage of avoiding the low efficiency that low Mach number shocks have for accelerating CR electrons out of the thermal pool. Before electrons can be efficiently accelerated at the shock, e.g., they must have gyro-radii large enough to carry them back and forth across the shock front. Whether mechanisms such as shock drift acceleration can be effective, or whether pre-existing seed populations are required, is yet to be determined. Pre-existing electrons will change the radio spectral signatures, so better radio- and X-ray- observational diagnostics and a range of numerical experiments are the order of the day.

## THE CORE of the MATTER

Deep in the cores of "relaxed" clusters, another problem is brewing. In this high density region the X-ray radiative cooling times of the hot ICM are short enough that 100s to 1000s of solar masses of material should be forming new stars each year. Observations indicate that only 10s of solar masses are turning on each year, and newer observations reveal cool molecular material – all exciting, but far from enough to meet the expectations.

Another view of the same issue is provided by the trends of what cluster astronomers call "entropy," $K=T/n^{2/3}$, a quantity that is relatively straightforward both to measure as a function distance from the cluster center, and to interpret. While T and n are both regulated by global cluster parameters such as mass and size, they aren't straightforward indicators of the energy flows. But when combined in this entropy form they are sensitive to the heating and cooling history of the ICM. In the case of successive accretion onto a cluster of increasing mass, the expected entropy profile is $K(r)=K_0 r^{1.1}$, close to what is observed beyond the cluster cores.

In the cores, the entropy drops more slowly, requiring some input of heat to the ICM, as shown in Figure 7. Ejections of relativistic plasma from the AGN of the bright central galaxy are one key source of this energy. Figure 8 shows the synchrotron lobes from a radio galaxy having excavated enormous cavities in its surrounding ICM. Episodic injections of up to $10^{55}$ J are observed, sometimes near the cluster core, and sometimes at much larger distances. The work done by such outbursts, as well as CR heating and sloshing motions appears sufficient to balance the cooling. This opens up the exciting possibility that clusters are self-regulating over cosmic times. In this scenario, the infall of material onto the central supermassive black hole is controlled by the cooling and instabilities near the cluster core which are themselves controlled by the energy released from the AGN outflows.

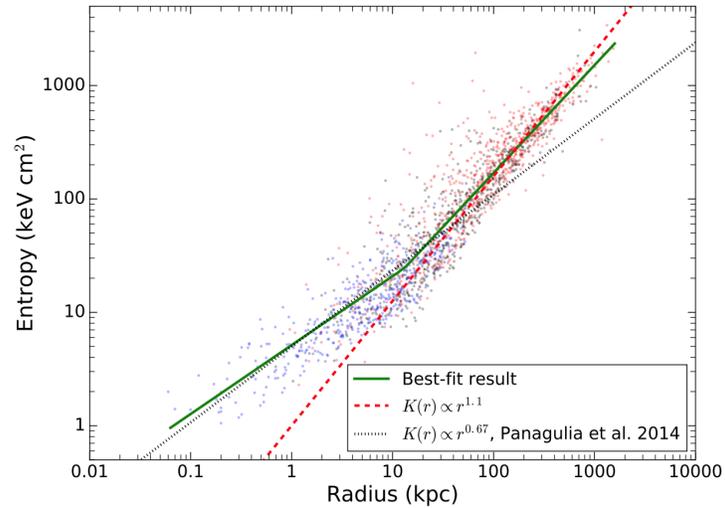

Figure 7. Entropy profiles become shallower in inner cluster regions, mostly due to AGN heating [9].

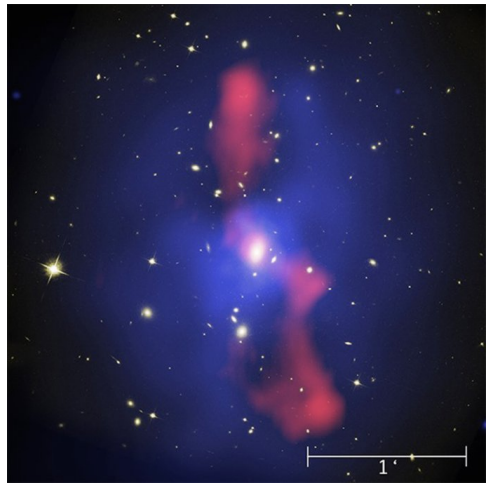

Figure 8. The inner 700 kpc of the MS0735.6+7421 cluster combining the X-ray (blue), *i*-band (white), and radio wavelengths (red) [10]

**THE WEATHER FORECAST**

In its far too brief life, the Japanese X-ray satellite Hitomi gave us a taste of things to come. With its exquisite spectral resolution, it directly measured the Doppler line broadening and centroid shifts of the lines from H- and He-like ions of Fe in the core of the Perseus cluster. The observed velocities were of order 100-200 km/s, providing the first direct measurements of the streaming and random gas motions in the central regions of a cluster ICM, due to both gravitational and AGN energy inputs. They also confirm that the ICM is still very close to hydrostatic equilibrium despite these exciting signposts of ICM activity.

Astronomers are quite excited about what's coming next. On the X-ray side, eROSITA, a joint German-Russian venture is projected to detect the ICM in 50-100 thousand clusters, dramatically expanding our capacity for statistical and cosmological studies. The exquisite spectral resolution and sensitivity of XRISM, a joint Japanese/U.S. mission, will map out the details of velocity structures in the ICM. Further down the road, the European satellite ATHENA will allow us to study the history of how the ICM is heated and chemically evolves over cosmic times, while NASA is considering the Lynx mission, which would be able to probe the diffuse medium in cluster-feeding filaments down to the astonishing low value of ~7 atoms/m$^3$. Complementing these are a new generation of radio telescopes and surveys which will probe the cosmic rays and magnetic fields. At low frequencies, the LOFAR Two Meter Survey and the Murchison Widefield Array have begun to reveal the extensive cluster synchrotron structures that have faded at higher frequencies due to radiative losses. Polarization at low frequencies, including new capabilities at the VLA, will provide detailed probes of magnetic field irregularities in the ICM. And large sky coverage polarization surveys such as the Polarization Sky Survey of the Universe's Magnetism on the Australian SKA Pathfinder telescope and the Very Large Array Sky Survey will allow statistical studies to connect the magnetic structures in the ICM with other cluster physical properties. Even more is on the horizon.

So clusters will continue to evolve, our instrumentation will get ever more powerful, our simulations increasingly sophisticated, and the weather outlook for the ICM is stormy indeed!

*Thanks to many colleagues across the spectrum, for their most wise counsel, and for support from the NSF through grant AST17-14205. And apologies to everyone for the lack of references, as per the PT style.*